\documentclass[prl,twocolumn,floatfix,showpacs]{revtex4}

\usepackage{graphicx}

\begin{document}

\title{Stochastic volatility of financial markets as the fluctuating
rate of trading: \\ an empirical study}

\author{A. Christian Silva}

\altaffiliation{Now at EvA Inc., San Francisco, silvaac@evafunds.com.}

\affiliation{Department of Physics, University of Maryland, 
College Park, MD 20742--4111, USA}

\author{Victor M. Yakovenko}

\affiliation{Department of Physics, University of Maryland, 
College Park, MD 20742--4111, USA}

\date{\bf physics/0608299, v.2 December 10, 2006}


\begin{abstract}
We present an empirical study of the subordination hypothesis for a
stochastic time series of a stock price.  The fluctuating rate of
trading is identified with the stochastic variance of the stock price,
as in the continuous-time random walk (CTRW) framework.  The
probability distribution of the stock price changes (log-returns) for
a given number of trades $N$ is found to be approximately Gaussian.
The probability distribution of $N$ for a given time interval $\Delta
t$ is non-Poissonian and has an exponential tail for large $N$ and a
sharp cutoff for small $N$.  Combining these two distributions
produces a nontrivial distribution of log-returns for a given time
interval $\Delta t$, which has exponential tails and a Gaussian
central part, in agreement with empirical observations.
\end{abstract}


\maketitle

\paragraph{Introduction: stochastic volatility, subordination, and 
fluctuations in the number of trades.}
\label{Sec:Subordination}

The stock price $S_t$ is a stochastic series in time $t$.  It is
commonly characterized by the probability distribution $P_{\Delta
t}(x)$ of detrended log-returns $x=\ln(S_{t_2}/S_{t_1})-\mu\Delta t$,
where the time interval $\Delta t=t_2-t_1$ is called the time lag or
time horizon, and $\mu$ is the average growth rate.  For a simple
multiplicative (geometric) random walk, the probability distribution
is Gaussian: $P_{\Delta t}(x)\propto\exp(-x^2/2v\Delta t)$, where
$v=\sigma^2$ is the variance, and $\sigma$ is the volatility.
However, the empirically observed probability distribution of
log-returns is not Gaussian.  It is well known that the distribution
has power-law tails for large $x$ \cite{Stanley99b,Stanley99a}.
However, the distribution is also non-Gaussian for small and moderate
$x$, where it follows the tent-shaped exponential (also called
double-exponential) Laplace law: $P_{\Delta
t}(x)\propto\exp(-c|x|/\sqrt{\Delta t})$, as emphasized in Ref.\
\cite{Silva04}.  The exponential distribution was found by many
researchers
\cite{Bouchaud,Miranda01,McCauley03,Kaizoji04,Remer04,Makowiec04,Matia04,Vicente06},
so it should be treated as a ubiquitous stylized fact for financial
markets \cite{Silva04}.

In order to explain the non-Gaussian character of the distribution of
returns, models with stochastic volatility were proposed in literature
\cite{Praetz72,Hull87,Heston93,Sircar-book}.  If the variance $v_t$
changes in time, then $v\Delta t$ in the Gaussian distribution should
be replaced by the integrated variance $V_{\Delta
t}=\int_{t_1}^{t_2}v_t\,dt$.  If the variance is stochastic, then we
should average over the probability distribution $Q_{\Delta t}(V)$ of
the integrated variance $V$ for the time interval $\Delta t$:
\begin{equation}
  P_{\Delta t}(x)=\int_0^\infty dV\, \frac{e^{-x^2/2V}}{\sqrt{2\pi V}}
  \, Q_{\Delta t}(V).
\label{V-subord}
\end{equation}
The representation (\ref{V-subord}) is called the subordination
\cite{Feller-book,Clark73}.  In this approach, the non-Gaussian
character of $P_{\Delta t}(x)$ results from a non-trivial distribution
$Q_{\Delta t}(V)$.

In the models with stochastic volatility, the variables $v$ or $V$ are
treated as hidden stochastic variables.  One may try to identify these
phenomenological variables with some empirically observable and
measurable components of the financial data.  It was argued
\cite{Mandelbrot67,Ane00,Stanley00} that the integrated variance
$V_{\Delta t}$ may correspond to the number of trades (transactions)
$N_{\Delta t}$ during the time interval $\Delta t$: $V_{\Delta t}=\xi
N_{\Delta t}$, where $\xi$ is a coefficient \cite{volume}.  Every
transaction may change the price up or down, so the probability
distribution $P_N(x)$ after $N$ trades would be Gaussian:
\begin{equation}
  P_N(x)=\frac{e^{-x^2/2\xi N}}{\sqrt{2\pi\xi N}}.
\label{P_N(x)}
\end{equation}
Then, the subordinated representation (\ref{V-subord}) becomes
\begin{equation}
  P_{\Delta t}(x)=\int_0^\infty dN\,
  \frac{e^{-x^2/2\xi N}}{\sqrt{2\pi\xi N}} \, K_{\Delta t}(N),
\label{N-subord}
\end{equation}
where $K_{\Delta t}(N)$ is the probability to have $N$ trades during
the time interval $\Delta t$.  (We assume that $N$ is large and use
integration, rather than summartion, over $N$.)  In this approach, the
stochastic variance $v$ reflects the fluctuating rate of trading in
the market.

Performing the Fourier transform of (\ref{N-subord}) with respect to
$x$, we find that the characteristic function $\tilde P_{\Delta
t}(k_x)$ is directly related to the Laplace transform $\tilde
K_{\Delta t}(k_N)$ of $K_{\Delta t}(N)$ with respect to $N$, where
$k_x$ and $k_N$ are the Fourier and Laplace variables conjugated to
$x$ and $N$:
\begin{equation}
  \tilde P_{\Delta t}(k_x) =\int\limits_0^\infty dN e^{-N\xi k_x^2/2}
  K_{\Delta t}(N) =\tilde K_{\Delta t}(\xi k_x^2/2).
\label{Fourier}
\end{equation}

In this paper, we study whether the subordinated representation
(\ref{N-subord}) agrees with financial data.  First, we verify whether
$P_N(x)$ is Gaussian, as suggested by Eq.\ (\ref{P_N(x)}).  Second, we
check whether empirical data satisfy Eq.\ (\ref{Fourier}).  Third, we
obtain $K_{\Delta t}(N)$ empirically and, finally, discuss whether
$P_{\Delta t}(x)$ constructed from Eq.\ (\ref{N-subord}) agrees with
the data.  Refs.\ \cite{Ane00,Stanley00} have already presented
evidence in favor of the first conjecture; however, the other
questions were not studied systematically in literature.

The subordination was also studied in physics literature as the
continuous-time random walk (CTRW) \cite{Montroll65,Montroll84}.
Refs.\ \cite{Scalas00,Sabatelli02,Masoliver03} focused on the
probability distribution of the waiting time $\Delta t$ between two
consequtive transactions ($\Delta N=1$).  Our approach is to study the
distribution function $K_{\Delta t}(N)$, which gives complementary
information and can be examined for a wide variety of time lags.  In
Ref.\ \cite{Dremin05}, this function was studied for some Russian
stocks.

We use the TAQ database from NYSE \cite{TAQ}, which records every
transaction in the market (tick-by-tick data).  We focus on the Intel
stock (INTC), because it is highly traded, with the average number of
transactions per day about $2.5\times10^4$.  Here we present the data
for the period 1 January -- 31 December 1999, but we found similar
results for 1997 as well \cite{Silva-thesis}.  Because of difficulties
in dealing with overnight price changes, we limit our consideration to
the intraday data.  Since $\Delta t$ is relatively short here, the
term $\mu\Delta t$ is small and can be neglected.

\paragraph{Probability distribution of log-returns $x$ after $N$ trades.}
\label{Sec:Gaussian}

\begin{figure}
\includegraphics[width=0.69\linewidth,angle=-90]{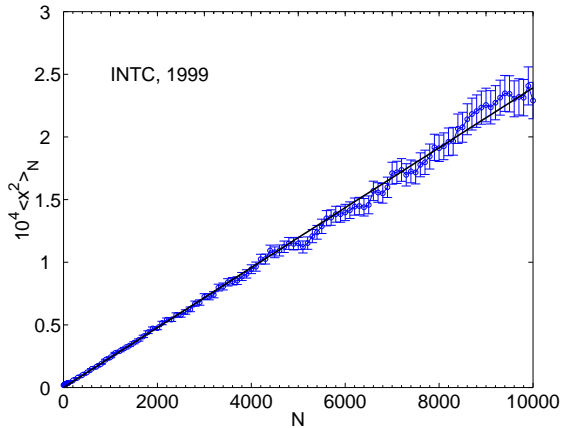}
\caption{The variance of log-returns $\langle x^2\rangle_N$ after $N$
  trades plotted vs.\ $N$.}
\label{fig:x^2N}
\end{figure}

\begin{figure}[b]
\includegraphics[width=0.69\linewidth,angle=-90]{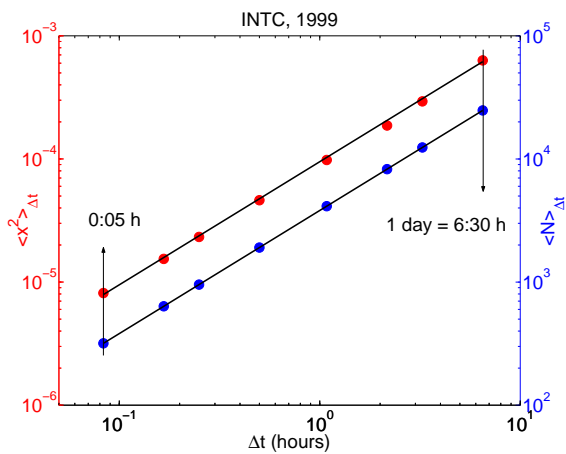}
\caption{The variance of log-returns $\langle x^2\rangle_{\Delta t}$
(upper points, left scale) and the average number of trades $\langle
N\rangle_{\Delta t}$ (lower points, right scale) vs.\ the time lag
$\Delta t$.  The solid lines of slope 1 represent the proportionality
relations (\ref{moments}).}
\label{fig:Nx^2t}
\end{figure}

\begin{figure}
\includegraphics[width=0.7\linewidth,angle=-90]{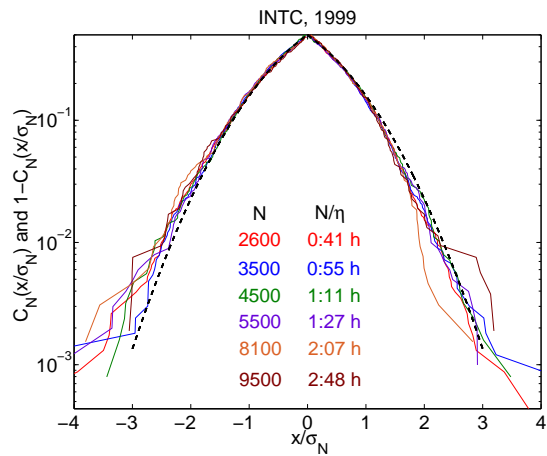}
\caption{Cumulative distribution $C_N(x/\sigma_N)$ of normalized
  log-returns after $N$ trades, where $\sigma_N^2=\langle
  x^2\rangle_N$, compared with the Gaussian distribution (dashed
  curve).  $N/\eta$ is the typical time interval between $N$ trades.}
\label{fig:CDF-xN}
\end{figure}

\begin{figure}[b]
\includegraphics[width=0.7\linewidth,angle=-90]{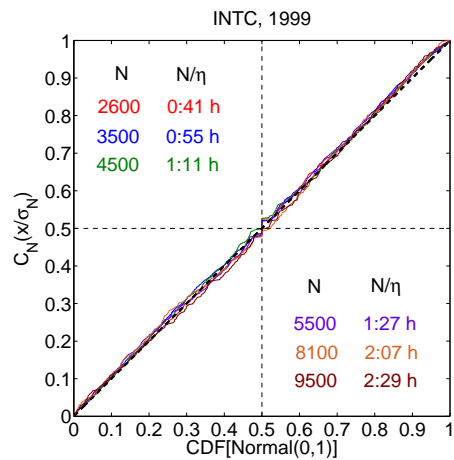}
\caption{The parametric Q-Q plot of the empirical cumulative
  distribution $C_N(x/\sigma_N)$ of normalized log-returns vs.\ the
  Gaussian distribution for the parameter $x$ from $-\infty$ to
  $+\infty$.}
\label{fig:QQ-xN}
\end{figure}

It follows from Eq.\ (\ref{P_N(x)}) that $\langle x^2\rangle_N=\xi N$,
where $\langle x^2\rangle_N$ is the second moment of $x$ after $N$
trades.  It is also natural to expect that the average number of
trades $\langle N\rangle_{\Delta t}$ during the time interval $\Delta
t$ is proportional to $\Delta t$ with some coefficient $\eta$.  Thus,
we expect
\begin{equation}
  \langle x^2\rangle_N=\xi N, \;
  \langle N\rangle_{\Delta t}=\eta\Delta t, \;
  \langle x^2\rangle_{\Delta t}=\theta\Delta t, \;
  \theta=\xi\eta.
\label{moments}
\end{equation}
Notice that the coefficient $\theta=\langle v\rangle$ is the mean
variance.  Figs.~\ref{fig:x^2N} and \ref{fig:Nx^2t} show that the
relations (\ref{moments}) are indeed satisfied.  We extract the values
of the coefficients from the slopes of these plots:
$\xi=2.4\times10^{-8}$ per one trade, $\eta=3.8\times10^3$
trades/hour, and $\theta=9.5\times10^{-5}$ per hour.  The relation
$\theta=\xi\eta$ is satisfied only approximately, but within the
measurement accuracy.

In Figs.\ \ref{fig:CDF-xN} and \ref{fig:QQ-xN}, we examine the
empirical probability distribution $P_N(x)$ of log-returns $x$ after
$N$ trades.  In Fig.~\ref{fig:CDF-xN}, the cumulative distribution
functions $C_N(x)=\int_{-\infty}^x dx'P_N(x')$ for $x<0$ and
$1-C_N(x)$ for $x<0$ are compared with the Gaussian distribution shown
by the dashed line.  The log-return $x$ is normalized by
$\sigma_N=\sqrt{\langle x^2\rangle_N}$.  The empirical distributions
$P_N(x)$ for different $N$ agree with the Gaussian in the central
part, but there are deviations in the tails, as expected for large
$|x|$.  Similar results were found in Fig.\ 6 of Ref.\
\cite{Farmer05}.

Fig.~\ref{fig:QQ-xN} shows the Q-Q plot similar to the one constructed
in Ref.\ \cite{Ane00}.  This is a parametric plot, where the vertical
axis shows the empirical $C_N(x/\sigma_N)$, and the horizontal axis
shows the cumulative Gaussian distribution of $x/\sigma$, whereas the
parameter $x$ changes from $-\infty$ to $+\infty$.  The plots for
different $N$ are all close to the diagonal, which indicates agreement
between the empirical and the Gaussian distribution functions.  Fig.\
\ref{fig:QQ-xN} emphasizes the central part of the distribution,
whereas Fig.~\ref{fig:CDF-xN} emphasizes the tails.  Overall, we
conclude the empirical distribution $P_N(x)$ is reasonably close to
the Gaussian in the central part, so Eq.\ (\ref{P_N(x)}) is
approximately satisfied.

When the time lag approaches one day, the number of data points become
too small to construct reliable probability densities, so we cannot
verify the Gaussian hypothesis beyond the intraday data.  When the
time lag is too short, and the corresponding $N$ is small, the
log-returns are discrete and cannot be described by a continuous
function, such as Gaussian.  We found that the distribution of $x$
becomes reasonably smooth only after a thousand of trades
\cite{Silva-thesis}.  Discreteness of the distribution for small $N$
can be seen in Fig.\ 11 of Ref.\ \cite{Farmer04}.

\paragraph{The characteristic function for log-returns and the Laplace 
transform for the number of trades.}
\label{Sec:Fourier}

\begin{figure}
\includegraphics[width=0.7\linewidth,angle=-90]{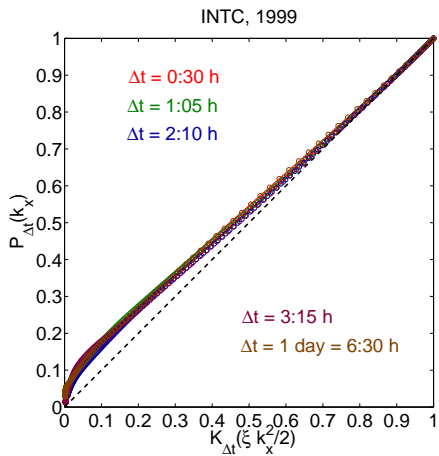}
\caption{The parametric plot of the Fourier transform $\tilde
  P_{\Delta t}(k_x)$ vs.\ the Laplace transform $\tilde K_{\Delta
  t}(\xi k_x^2/2)$ verifying Eq.\ (\ref{Fourier}) for the parameter
  $k_x$ from $-\infty$ to $+\infty$.}
\label{fig:Fourier}
\end{figure}

The subordination hypothesis (\ref{N-subord}) can be examined further
by checking the relation (\ref{Fourier}) between the Fourier transform
$\tilde P_{\Delta t}(k_x)$ for log-returns and the Laplace transform
$\tilde K_{\Delta t}(k_N)$ for the number of trades.  These functions
can be directly constructed from the data.  As shown in Ref.\
\cite{Silva04}, $\tilde P_{\Delta
t}(k_x)\approx(1/n)\sum_{j=1}^ne^{ik_xx_j}$ and $\tilde K_{\Delta
t}(k_N)\approx(1/m)\sum_{l=1}^me^{-k_NN_l}$, where the sums are taken
over all occurrences of the log-returns $x_j$ and the numbers of
trades $N_l$ during a time interval $\Delta t$ in a dataset.  Because
the frequency of appearances of a given $x_j$ or $N_l$ is proportional
to the corresponding probability density, these sums approximate the
integral definitions $\tilde P_{\Delta t}(k_x)=
\int_{-\infty}^{+\infty}dx\,e^{ik_xx}P_{\Delta t}(x)$ and $\tilde
K_{\Delta t}(k_N)= \int_{0}^{\infty}dN\,e^{-k_NN}K_{\Delta t}(N)$.

In Fig.~\ref{fig:Fourier}, we show the parametric plot of $\tilde
P_{\Delta t}(k_x)$ vs.\ $\tilde K_{\Delta t}(\xi k_x^2/2)$.  The
vertical axis shows $\tilde P_{\Delta t}(k_x)$, and the horizontal
axis shows $\tilde K_{\Delta t}(\xi k_x^2/2)$, whereas the parameter
$k_x$ changes from from $-\infty$ to $+\infty$.  The upper right
corner $(1,1)$ corresponds to $k_x=0$, and the lower left corner
$(0,0)$ corresponds to large $|k_x|$.  The parameter $\xi$ used in Fig.\
\ref{fig:Fourier} is extracted from the slope of $\langle
x^2\rangle_N$ vs.\ $N$ in Fig.~\ref{fig:x^2N}.  The relations
\begin{equation}
  \left.\frac{d^2\tilde P_{\Delta t}(k_x)}{dk_x^2}\right|_{k_x=0}
  =-\langle x^2\rangle_{\Delta t}, \:
  \left.\frac{d\tilde K_{\Delta t}(k_N)}{dk_N}\right|_{k_N=0}
  =-\langle N\rangle_{\Delta t}
\label{moments-Fourier}
\end{equation}
and Eq.\ (\ref{moments}) ensure that the slope of the parametric plot
near the point $(1,1)$ corresponds to the diagonal.  Overall, the
plots for different $\Delta t$ in Fig.~\ref{fig:Fourier} are close to
the diagonal, but deviate in the lower corner for large $|k_x|$, which
indicates that the subordination relation (\ref{Fourier}) is satisfied
only approximately.  Notice that no assumptions about the functional
form of $K_{\Delta t}(N)$ are made in Eq.\ (\ref{Fourier}).  The only
assumption is that $P_N(x)$ is Gaussian (\ref{P_N(x)}), and the
distributions of $x$ and $N$ are uncorrelated, so they can be combined
in Eq.\ (\ref{N-subord}).

\paragraph{Probability distribution of the number of trades $N$ during
  the time interval $\Delta t$.}
\label{Sec:Nt}

\begin{figure}
\includegraphics[width=0.72\linewidth,angle=-90]{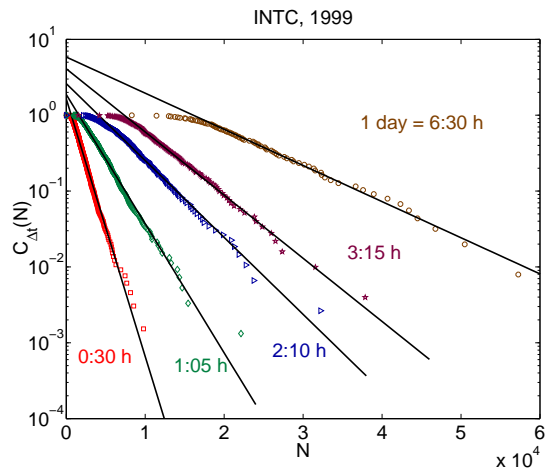}
\caption{Empirical cumulative distribution $C_{\Delta t}(N)$ for the
  number of trades $N$ during the time interval $\Delta t$, plotted in
  the log-linear scale.  The solid lines illustrate exponential
  tails.}
\label{fig:cdfNt}
\end{figure}

Fig.~\ref{fig:cdfNt} shows the log-linear plot of the empirically
constructed cumulative distribution $C_{\Delta t}(N)=\int_N^\infty
K_{\Delta t}(N')\,dN'$ for the number of trades $N$ during the time
interval $\Delta t$.  The straight lines are eye guides, which
indicate that the probability distributions $K_{\Delta t}(N)$ are
exponential for large $N$.  The slopes of the lines are related to
$\langle N\rangle_{\Delta t}=\eta\Delta t$, so we can approximate
$K_{\Delta t}(N)\propto\exp(-N/\eta\Delta t)$ for large $N$.  For
small $N$, Fig.~\ref{fig:cdfNt} shows that $C_{\Delta t}(N)$ is flat
and $K_{\Delta t}(N)$ is suppressed, so that $K_{\Delta t}(N=0)=0$.
It is indeed very improbable to have no trades at all for an extended
time period $\Delta t$.  For long enough $\Delta t$, we expect that
$K_{\Delta t}(N)$ would become a Gaussian function of $N$ centered at
$\langle N\rangle_{\Delta t}=\eta\Delta t$.  However, this regime has
not been achieved yet for the time lags $\Delta t$ shown in Fig.\
\ref{fig:cdfNt}.  For short $\Delta t$, we also found that $\langle
N^2\rangle_{\Delta t}\propto\langle N\rangle^2_{\Delta t}$ with a
coefficient somewhat smaller than 2, as expected for an approximately
exponential distribution.

Notice that the exponential behavior of the empirical $K_{\Delta
t}(N)$ shown in Fig.\ \ref{fig:cdfNt} is inconsistent with the Poisson
distribution $K_{\Delta t}^{\rm Poisson}(N)=e^{-\eta\Delta
t}(\eta\Delta t)^N/N!$ expected for trades occurring randomly and
independently at the average rate $\eta$.  It was suggested in
literature that $K_{\Delta t}(N)$ may be approximated by the
log-normal or gamma distributions.  We do not attempt to discriminate
between the alternative hypotheses here, but sometimes these functions
may look alike \cite{Banerjee06}.  A qualitatively similar
distribution $K_{\Delta t}(N)$ was found for some Russian stocks in
Ref.\ \cite{Dremin05}.

\paragraph{Probability distribution of log-returns $x$ after
  the time interval $\Delta t$.}
\label{Sec:xt}

\begin{figure}
\includegraphics[width=0.8\linewidth,angle=-90]{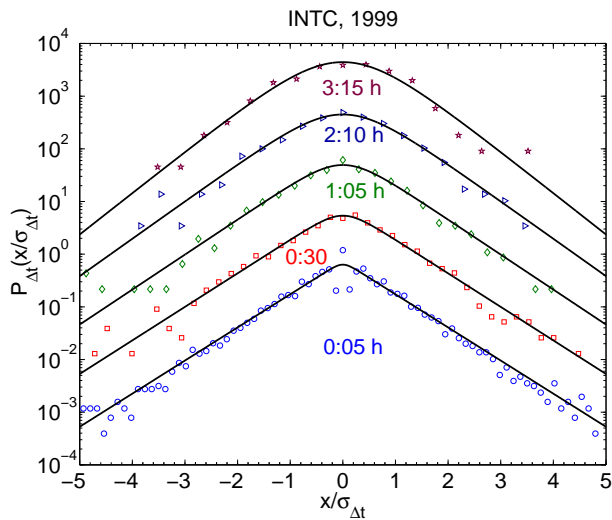}
\caption{Probability density $P_{\Delta t}(x/\sigma_{\Delta t})$ of
  normalized log-returns after the time lag $\Delta t$, where
  $\sigma_{\Delta t}^2=\langle x^2\rangle_{\Delta t}$.  The solid
  lines are fits to the Heston model with $1/\gamma=50$ min.  The
  curves are offset vertically by the factors of 10.}
\label{fig:pdfXt}
\end{figure}

Having established that $P_N(x)$ is approximately Gaussian
(\ref{P_N(x)}), and $K_{\Delta t}(N)$ is approximately exponential for
short $\Delta t$, we can obtain $P_{\Delta t}(x)$ from Eq.\
(\ref{N-subord}).  Substituting these expressions into Eq.\
(\ref{N-subord}), we get
\begin{equation}
  P_{\Delta t}(x)\approx\int_0^\infty dN \, 
  \frac{e^{-x^2/2\xi N}}{\sqrt{2\pi\xi N}} \, 
  \frac{e^{-N/\eta\Delta t}}{\eta\Delta t}
  =\frac{e^{-|x|\sqrt{2/\theta\Delta t}}}{\sqrt{2\theta\Delta t}}.
\label{exp}
\end{equation}
Eq.\ (\ref{exp}) shows that the exponential distribution of the number
of trades $N$ results in the exponential (Laplace) distribution of
log-returns $x$.  This can be understood as follows.  The integral
(\ref{exp}) can be taken exactly, but one can also evaluate it
approximately by integrating around the optimal value of
$N_*=|x|\sqrt{\eta\Delta t/2\xi}$ that minimizes the negative
expression in the exponent of Eq.\ (\ref{exp}) and maximizes the
integrand.  We see that the probability to have a given log-return $x$
is controlled by the probability to have the optimal number of trades
$N_*(x)$.  Thus, the distribution $P_{\Delta t}(x)$ has the fatter
(exponential) tails than Gaussian, because the probability to have a
large $x$ is enhanced by fluctuations with large $N$.

On the other hand, for very small $x$, the optimal value $N_*$ becomes
limited by the cutoff in $K_{\Delta t}(N)$ for small $N$.  At this
point, the optimal value $N_*$ stops depending on $x$, so $P_{\Delta
t}(x)$ becomes Gaussian.  Thus, we expect to see the Gaussian behavior
in $P_{\Delta t}(x)$ for small $|x|$ and the exponential behavior for
medium and large $|x|$.  Fig.~\ref{fig:pdfXt} shows a log-linear plot
of the empirical probability density $P_{\Delta t}(x)$.  In agreement
with the qualitative analysis presented above, we observe that the
data points follow the parabolic (Gaussian) curve for small $|x|$ and
fall on the straight (exponential) lines for large $|x|$.  The range
of $x$ occupied by the Gaussian expands when the time lag $\Delta t$
increases, because the cutoff in $K_{\Delta t}(N)$ for small $N$
increases with the increase of $\Delta t$, as shown in
Fig.~\ref{fig:cdfNt}.  We conclude that the subordination hypothesis
(\ref{N-subord}) is qualitatively valid, and, particularly, it
explains the exponential distribution $P_{\Delta t}(x)$ for $x$ as a
result of the exponential distribution $K_{\Delta t}(N)$ for the
number of trades $N$.

The solid lines in Fig.~\ref{fig:pdfXt} show fits of the data to the
Heston model.  The Heston model \cite{Heston93} is a model with
stochastic volatility, which has the advantage of being exactly
solvable.  A closed-form solution for $P_{\Delta t}(x)$ was obtained
in Ref.\ \cite{Dragulescu02}, and Fig.~\ref{fig:pdfXt} shows fits of
the data to the formula derived there.  Refs.\
\cite{Silva04,Dragulescu02} pointed out that $P_{\Delta t}(x)$ in the
Heston model has the exponential tails and Gaussian center, in
qualitative and quantitative agreement with the empirical distribution
of log-returns.  Given the verification of the subordination
hypothesis presented in this paper, one may ask whether the Heston
model describes the probability distribution $K_{\Delta t}(N)$ for the
number of trades $N$.  A detailed study of this question will be
presented in a separate paper \cite{U-shape}.

We also would like to point out that Eq.\ (\ref{exp}) represents a
special case of the variance-gamma distribution introduced by Madan
and Seneta \cite{Madan90}.  The Heston model solution
\cite{Dragulescu02} reduces to the variance-gamma distribution in the
limit of short $\Delta t$, see Eqs.\ (48) and (49) in Ref.\
\cite{Dragulescu02}.




\begin{thebibliography}{99}
\bibitem{volume} One may also consider the volume of trades instead of
  the number of trades.  In that case, empirical analysis is
  technically more complicated, so we limit our consideration to the
  number of trades.
\bibitem{U-shape} We are also aware that the variation of volatility
  during a day has the well-known U-shape, with the highest number of
  trades at the opening and closing times \cite{Silva-thesis}.
\end{thebibliography}
\end{document}